\documentclass[journal=jacsat,manuscript=article]{achemsoz}
\usepackage[version=3]{mhchem} 
\usepackage{graphicx}
\usepackage{float}
\usepackage{subcaption}
\usepackage{color}
\usepackage{gensymb}
\usepackage{setspace}
\makeatletter
\author{Mohamed Zbiri}
\affiliation{Institut Laue-Langevin, 71 avenue des Martyrs, Grenoble Cedex 9, 38042, France}
\email{zbiri@ill.fr}

\author{Peter A. Finn}
\affiliation{Materials Research Institute and School of Biological and Chemical Sciences, Queen Mary University of London, Mile End Road, London E1 4NS, United Kingdom}

\author{Christian B. Nielsen}
\affiliation{Materials Research Institute and School of Biological and Chemical Sciences, Queen Mary University of London, Mile End Road, London E1 4NS, United Kingdom}

\author{Anne A.Y. Guilbert}
\affiliation{Department of Physics and Centre for Plastic Electronics, Imperial College London, Prince Consort Road, London SW7 2AZ, United Kingdom}
\email{a.guilbert09@imperial.ac.uk}
\makeatletter
\let\acs@address@list\relax
\makeatother
\title{Quantitative insights into the phase behaviour and miscibility of organic photovoltaic active layers from the perspective of neutron spectroscopy}
\begin{document}
\begin{abstract}
We present a neutron spectroscopy based method to study quantitatively the partial miscibility and phase behaviour of an organic photovoltaic active layer made of conjugated polymer:small molecule blends, presently illustrated with the regio-random poly(3-hexylthiophene-2,5-diyl) and fullerene [6,6]-Phenyl C$_{61}$ butyric acid methyl ester (RRa-P3HT:PCBM) system. We perform both inelastic neutron scattering and quasi-elastic neutron scattering measurements to study the structural dynamics of blends of different compositions enabling us to resolve the phase behaviour. The difference of neutron cross sections between RRa-P3HT and PCBM, and the use of deuteration technique, offer a unique opportunity to probe the miscibility limit of fullerene in the amorphous polymer-rich phase and to tune the contrast between the polymer and the fullerene phases, respectively. Therefore, the proposed approach should be universal and relevant to study new non-fullerene acceptors that are closely related - in terms of chemical structures - to the polymer, where other conventional imaging and spectroscopic techniques present a poor contrast between the blend components.
\end{abstract}
\section{Introduction}
Excitons, generated upon light absorption in conjugated polymers, are known to dissociate into free charges in the presence of an electron acceptor material. Bulk heterojunctions made of polymer donor and small-molecule acceptor materials constitute the active layer of organic solar cells. The microstructure of such blends is complex with most likely three phases, a small-molecule rich phase, an amorphous polymer-rich phase and if the polymer is semi-crystalline, a pure crystalline polymer phase.\cite{Collins2011} Only few polymers such as poly[2,5-bis(3-tetradecylthiophen-2-yl)thieno[3,2-b]thiophene] (PBTTT)\cite{Jamieson2012} are known to co-crystallise with fullerene acceptors. Because of the asymmetry of the donor and acceptor molecular weights, the small molecule rich phase is nearly pure. The amorphous polymer-rich phase is beneficial for charge separation\cite{Westacott2013} while the presence of nearly pure percolated donor and acceptor domains are beneficial for the transport of charges generated at the donor:acceptor heterojunction to the electrodes.\\
If crystallinity is relatively simple to monitor by methods such as X-Ray diffraction, the composition of the amorphous mixture of the blends\cite{Treat2011, Collins2013} as well as changes in conformation with respect to the neat materials is more difficult to access.\cite{Tsoi2011, Razzell2016} Although crystallinity has been shown to improve charge transport\cite{Noriega2013} and potentially lead to extra driving force for charge separation by lowering the electronic energy levels,\cite{Jamieson2012} a spinodal-type decomposition emerged as a new picture for phase separation at length scales directly relevant to the operation of the devices,\cite{Treat2014,Li2017,Ye2018} with the coarsening of this phase separation directly linked to burn in degradation mechanisms.\cite{Li2017} The crucial role of the Flory-Huggins interaction parameter ($\chi$) in controlling phase behaviour, i.e. miscibility in the amorphous phase has been emphasised and related to solar cell efficiency.\cite{Ye2018,Peng2018} $\chi$ is both composition- and temperature-dependent, and is related to the thermodynamical stability of the blend. However, the formation of the bulk heterojunction proceeds through solution processing.\cite{Richter2017} Thus, the final microstructure is not thermodynamically stable but  kinetically trapped. Crystal seeds of small molecules and more or less large crystals of the polymer may form in the solution depending on the quality of the solvent for each component of the blend.\cite{Troshin2009, Machui2011, Machui2012} Moreover, liquid-liquid demixing may occur during solvent evaporation which could contribute to enhance phase separation.\\
\begin{figure}[H]
\includegraphics[width=18cm]{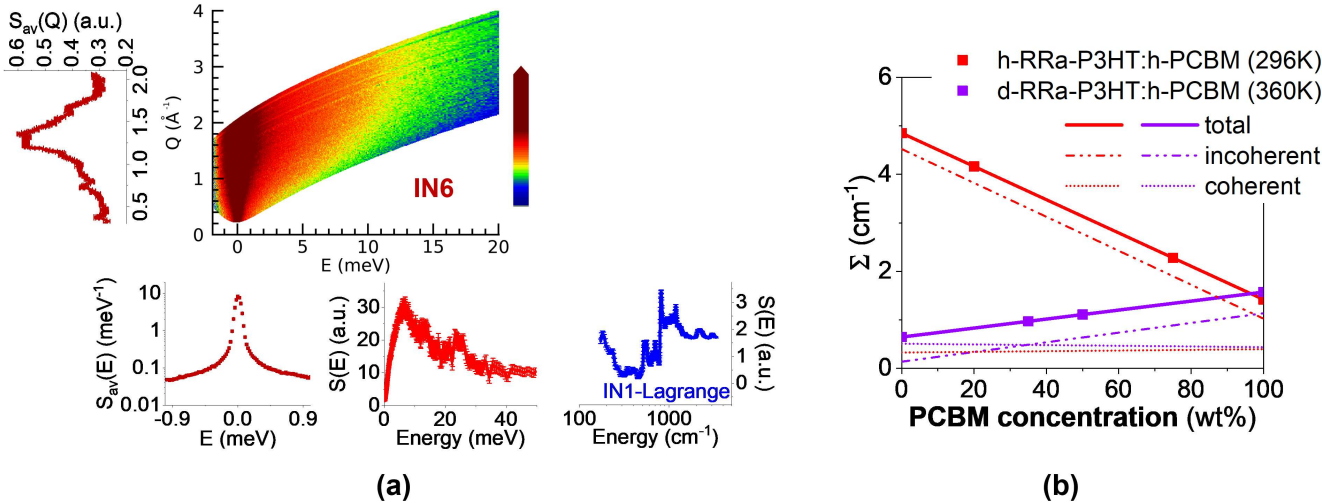}
\caption{(a) Schematic illustration of the quantities extracted from the neutron spectroscopic measurements. From the measured dynamical structure function, S(Q, E) (color coded map), at 360 K using IN6, the diffraction pattern (left), QENS spectrum, and low-energy INS spectrum (bottom) are obtained. The mid-to-high energy vibrational spectrum (bottom right) was measured at 10 K using IN1-Lagrange.  (b) Total, incoherent and coherent macroscopic neutron cross-sections ($\Sigma$) as a function of h-PCBM concentration in the presently studied samples. The samples represented by scatter points are h-RRa-P3HT:h-PCBM at 296 K (0 wt\%, 20 wt\%, 75 wt\% and 100 wt\% h-PCBM) and d-RRa-P3HT:h-PCBM at 360 K (0 wt\%, 35 wt\%, 50 wt\%, 100 wt\% h-PCBM). The macroscopic neutron cross-sections are extracted from the QENS data as explained in Supporting Information.} \label{fgr:fig1}
\end{figure}
Previously, we used a combination of quasi-elastic neutron scattering (QENS) measurements\cite{Guilbert2016} and molecular dynamics (MD) simulations\cite{Guilbert2017} to investigate the impact of each component of a blend of regio-regular poly(3-hexylthiophene-2,5-diyl) (RR-P3HT) and fullerene [6,6]-Phenyl C$_{61}$ butyric acid methyl ester  (PCBM) on their respective dynamics. We observed that, upon blending with PCBM, the QENS signal of P3HT is narrowing, while upon blending with RR-P3HT, the QENS signal of PCBM is broadening. We did interpret these observations as a signature of the frustration of RR-P3HT and plasticization of PCBM upon blending, respectively. The frustration of RR-P3HT was also observed by other groups on a different time scale. \cite{Paterno2013,Etampawala2015,Paterno2016} Our MD simulations suggested that these changes were due to the partial miscibility of P3HT:PCBM, in particular the formation of an amorphous mixture of P3HT:PCBM. MD simulations further revealed a conformational change of P3HT chain to accommodate PCBM with enhanced cofaciality between the polymer thiophene rings and the PCBM cage. This has further been supported by Zheng \textit{et al.}, whose MD simulations pointed towards the same cofaciality between P3HT and PCBM. They calculated the transfer integrals between P3HT and PCBM in such arrangement, \cite{Zheng2019} concluding that the enhanced cofaciality was beneficial for the charge separation processes in organic solar cells.\\
Presently, we study blends of regio-random P3HT and PCBM (RRa-P3HT:PCBM), with various compositions of PCBM, to clarify the impact of partial miscibility of conjugated polymer:small molecule systems. With respect to previous studies,\cite{Paterno2013,Etampawala2015,Guilbert2016,Paterno2016} we go a step further by using both inelastic neutron scattering (INS) and quasi-elastic neutron scattering (QENS) (Figure \ref{fgr:fig1} a) to resolve simultaneously changes in microstructure, morphology and dynamics of both the polymer and fullerene upon blending as a function of temperature. The observed morphological changes are further rationalised by quantum chemical calculations. To gain deeper insights, we use the deuteration technique to vary the contrast between the polymer and the fullerene. In the following, hydrogenated and deuterated materials will be labeled with the prefix h- and d-, respectively. The different sample compositions and their neutron cross sections are presented in Figure \ref{fgr:fig1} b. We propose to take advantage of the difference of neutron cross sections between conjugated polymer and fullerene to evaluate the miscibility limit of fullerene within the amorphous polymer-rich phase. The neutron spectroscopy based method presently described should be universal and relevant to study blends with new non-fullerene acceptors that are closely related in terms of chemical structures to the polymer, which otherwise lead to a poor contrast between the blend components when using conventional imaging and optical spectroscopy techniques.\cite{Rezasoltani2020}
\section{Results and discussion}
\subsection{Evaluating the phase composition of the blends}
\begin{figure}[H]
\includegraphics[width=18cm]{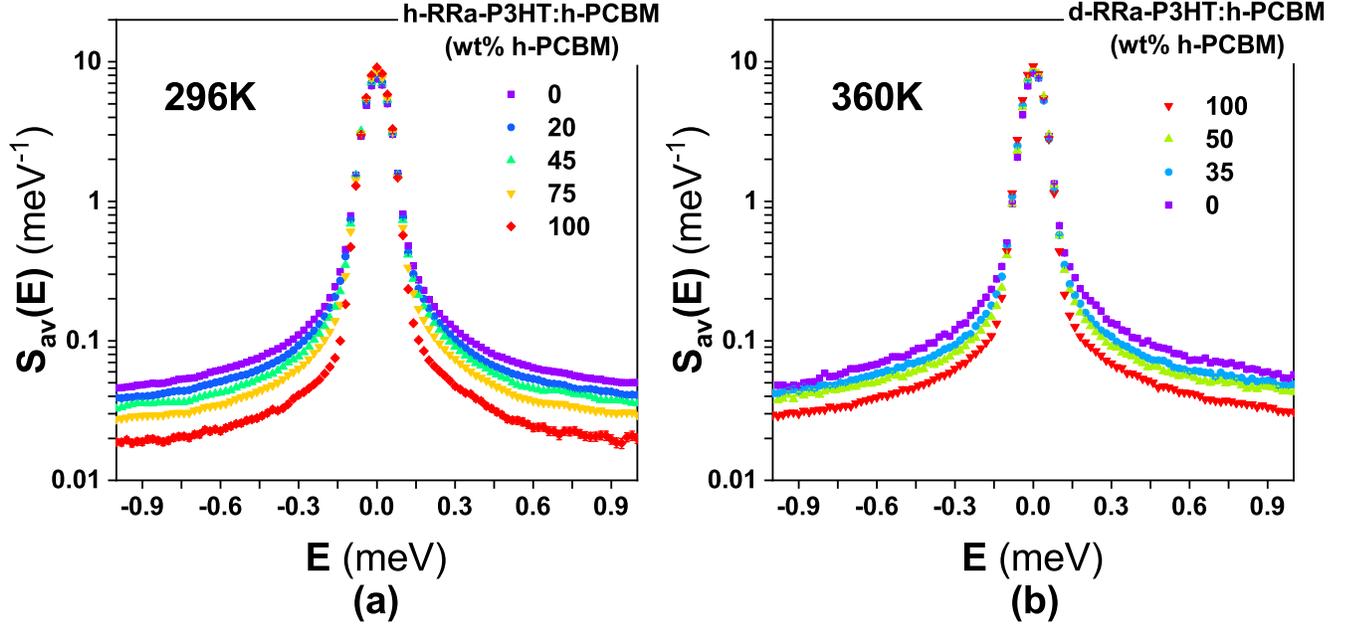}
\caption{Q-averaged QENS spectra of (a) neat h-RRa-P3HT, blends of h-RRa-P3HT:h-PCBM of 20 wt\%, 45 wt\% and 75 wt\% h-PCBM concentration and neat h-PCBM at 296K and (b) neat d-RRa-P3HT, blends of d-RRa-P3HT:h-PCBM of 35 wt\% and 50 wt\% h-PCBM concentration and neat h-PCBM at 360K.} \label{fgr:fig2}
\end{figure}

\begin{figure}[H]
\includegraphics[width=18cm]{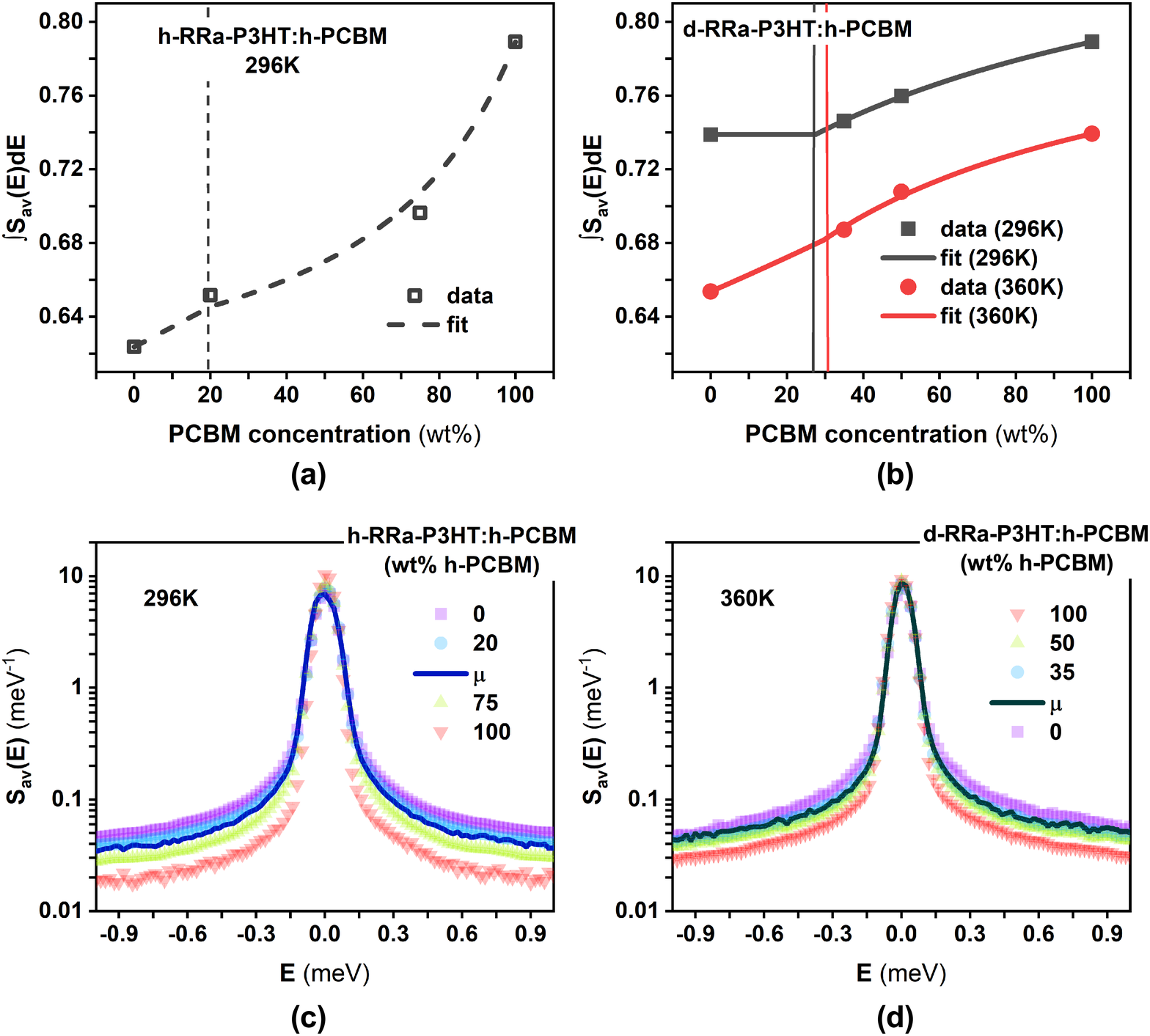}
\caption{ (a) and (b) Concentration-dependent integrals (scatter points) of the elastic region of the Q-averaged QENS spectra of h-RRa-P3HT:h-PCBM at 296K, and d-RRa-P3HT:h-PCBM at 296K and 360K, respectively. The integration was done between -0.05 and 0.05 meV. Lines are fits using logistic functions-based Equation \ref{eq:eq_3}. (c) Q-averaged QENS spectra of h-RRa-P3HT, h-RRa-P3HT:h-PCBM of 20 wt\%, $\mu$ ~ 20 wt\%, and 75 wt\% h-PCBM concentration and h-PCBM at 296K. (d) Q-averaged QENS spectra of d-RRa-P3HT, blends of d-RRa-P3HT:h-PCBM of $\mu$ ~ 30 wt\% ,35 wt\% and 50 wt\% h-PCBM concentration and h-PCBM at 360K. $\mu$ is the miscibility concentration for each system at the studied temperature.} \label{fgr:fig3}
\end{figure}

Figures \ref{fgr:fig2} a and b show the Q-averaged QENS spectra of neat h-RRa-P3HT, blends of h-RRa-P3HT:h-PCBM of compositions of 20 wt\%, 45 wt\% and 75 wt\% h-PCBM, neat h-PCBM at 296K, and neat h-PCBM, blends of d-RRa-P3HT:h-PCBM of compositions of 35 wt\% and 50 wt\% h-PCBM, neat d-RRa-P3HT at 360K, respectively. As the concentration of h-PCBM increases in the h-RRa-P3HT:h-PCBM blends, the QENS spectra are narrowing while as the concentration of d-RRa-P3HT increases in the blends of d-RRa-P3HT:h-PCBM, the QENS spectra are broadening. The incoherent contribution of h-PCBM is too small compared to that of h-RRa-P3HT, thus, the QENS signal is largely dominated by the h-RRa-P3HT incoherent contribution. However, as the concentration of h-PCBM increases, the associated contribution is expected to become more significant. We selected concentrations where the overall incoherent contribution to the signal is higher than the coherent counterpart,\cite{Paterno2016} although it is the difference which is reduced in the case of d-RRa-P3HT (Figure \ref{fgr:fig1} b). It is therefore reasonable to state that adding h-PCBM to h-RRa-P3HT frustrates the P3HT dynamics, and adding d-RRa-P3HT to h-PCBM plasticizes PCBM, in agreement with previous findings.\cite{Guilbert2016}\\
Having determined the macroscopic densities (see Supporting Information) of the neat polymer and fullerene phases, $\Sigma_{RRa-P3HT}$ and $\Sigma_{PCBM}$ (Figure \ref{fgr:fig1} b), respectively, we can proceed with modelling the QENS data to gain quantitative insights into the concentration-dependent phase behaviour, as shown in Figure \ref{fgr:fig3} a and b. RRa-P3HT is fully amorphous and thus, it is reasonable to assume that the studied samples with an overall PCBM concentration higher than the miscibility limit, $\mu$, exhibit two phases; a nearly pure h-PCBM phase and an amorphous RRa-P3HT rich phase. The QENS signal can, therefore, be expressed for an overall h-PCBM concentration $c_0$ larger than $\mu$ as follow:
\begin{multline}
(c_0 \times \Sigma_{PCBM} + (1-c_0) \times \Sigma_{RRa-P3HT}) \times S(c_0,E,Q) = \\
\frac{c_0-\mu}{1-\mu} \times \Sigma_{PCBM} \times S^{PCBM}(E,Q) + \\ \frac{1-c_0}{1-\mu} \times (\mu \times \Sigma_{PCBM} + (1-\mu) \times \Sigma_{RRa-P3HT}) \times S^{mix}(\mu,E,Q)
\label{eq:eq_1}
\end{multline}
where $S(c_0,E,Q)$ is the concentration-dependent total dynamical scattering function, $S^{PCBM}(E,Q)$ is the dynamical scattering function of the PCBM phase and $S^{mix}(\mu,E,Q)$ is the mixed-phase dynamical scattering function at the miscibility limit. Below the miscibility limit $\mu$, it is reasonable to assume that we have a solid solution (amorphous mixture) and we assume that:
\begin{equation}
\frac{\mu}{S^{mix}(c_0,E,Q)} = \frac{c_0}{S^{mix}(\mu,E,Q)} + \frac{\mu-c_0}{S^{RRa-P3HT}(E,Q)}
\label{eq:eq_2}
\end{equation}
where $S^{mix}(c_0,E,Q)$ is the concentration-dependent dynamical scattering function of the mixed-phase below the miscibility limit and $S^{RRa-P3HT}(E,Q)$ is the dynamical scattering function of the RRa-P3HT phase. In order to describe continuously the change and evolution in phase behaviour, we use logistic functions with a large $k$ parameter to approximate step functions, so the previous equation becomes:
\begin{multline}
S(c_0,E,Q) = \frac{1}{1+e^{k(c_0-\mu)}} \times {\frac{\mu}{\frac{c_0}{S^{mix}(\mu,E,Q)} + \frac{\mu-c_0}{S^{RRa-P3HT}(E,Q)}}} + \\
\frac{1}{1+e^{-k(c_0-\mu)}} \times \{ \frac{c_0-\mu}{1-\mu} \times \frac{\Sigma_{PCBM}}{c_0 \times \Sigma_{PCBM} + (1-c_0) \times \Sigma_{RRa-P3HT}} \times S^{PCBM}(E,Q)  \\ 
+  \frac{1-c_0}{1-\mu} \times \frac{\mu \times \Sigma_{PCBM} + (1-\mu) \times \Sigma_{RRa-P3HT}}{c_0 \times \Sigma_{PCBM} + (1-c_0) \times \Sigma_{RRa-P3HT}} \times S^{mix}(\mu,E,Q) \}
\label{eq:eq_3}
\end{multline}
The two quantities to fit are the miscibility $\mu$ and the scattering intensity at the miscibility limit $S^{mix}(\mu,E,Q)$. We fit successfully the QENS spectra using this model (Figures \ref{fgr:fig3} a and b). We found miscibility limits of about 20 wt\% PCBM for h-RRa-P3HT:h-PCBM at 296 K, 27 wt\% PCBM and 30 wt\% PCBM for d-RRa-P3HT:h-PCBM at 296 K and 360 K, respectively. These values are within the range found by means of other techniques.\cite{Yin2011,Treat2011,Collins2011,Guilbert2014} Furthermore, we found that as expected and supported by other techniques, the miscibility is increasing slightly with temperature.\cite{Collins2011} The observed difference in the miscibility limits obtained at 296K for h-RRa-P3HT:h-PCBM and d-RRa-P3HT:h-PCBM can be attributed to factors like deuteration, difference in molecular weight of the two polymers and the difference in regioregularity (Table \ref{tbl:tbl_1}). Interestingly, this simple model assuming two phases above the miscibility limit captures well the microstructure of the blends, hence allowing us to extract the QENS spectra at the miscibility limit for each blend at different temperatures (Figures \ref{fgr:fig3} c and d). The further narrowing of the QENS spectra of h-P3HT:h-PCBM for overall PCBM concentrations larger than the miscibility limit is not due to an extra frustration of the RRa-P3HT but to the presence of the almost neat crystalline PCBM phase.
\subsection{Monitoring simultaneously miscibility and microstructure}
\begin{figure}[H]
\includegraphics[width=18cm]{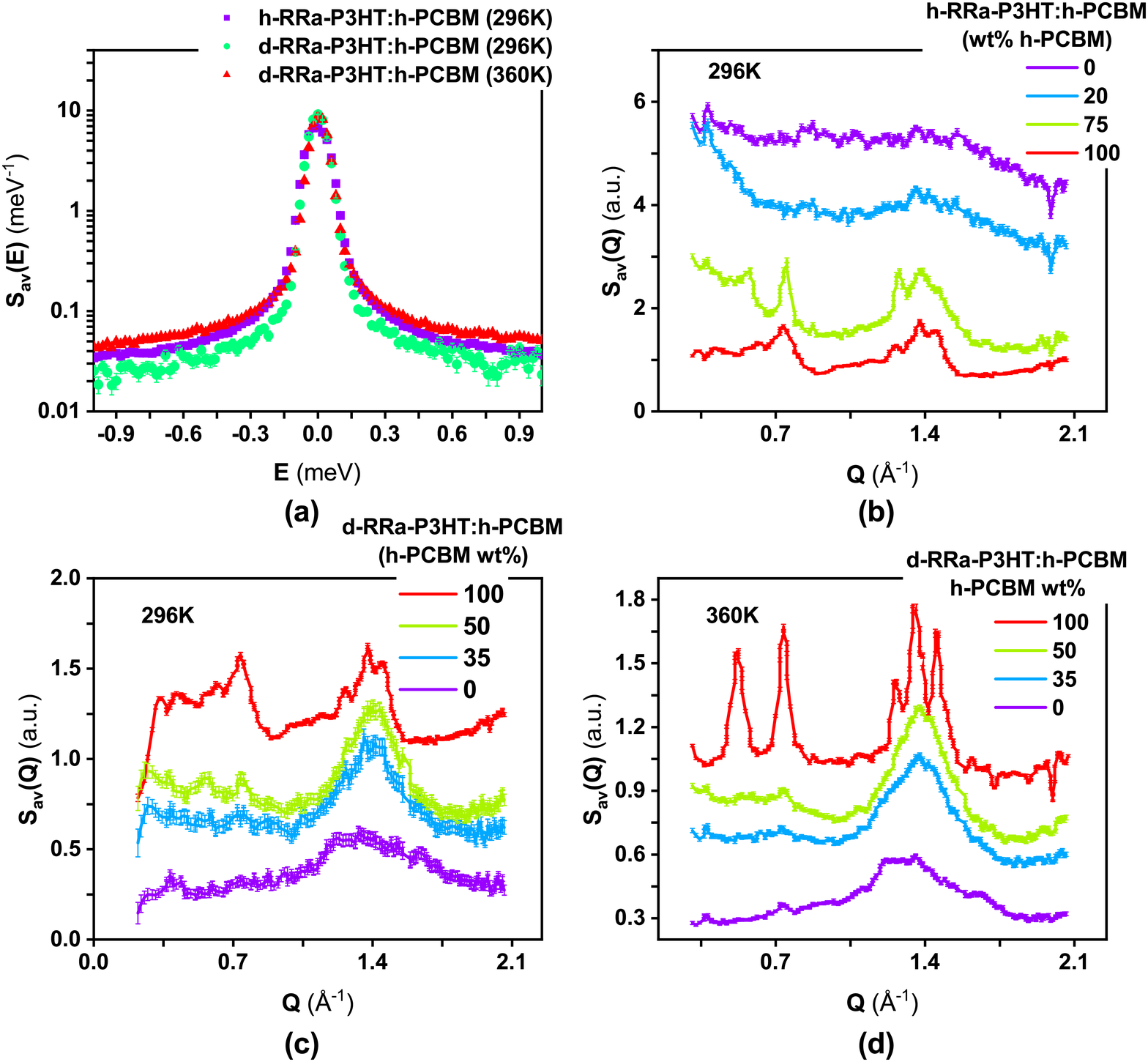}
\caption{(a) Q-averaged QENS spectra at the miscibility limit. (b) concentration-dependent neutron diffraction patterns of h-RRa-P3HT:h-PCBM at 296K. (c) and (d ) concentration-dependent neutron diffraction patterns of d-RRa-P3HT:h-PCBM at 296K and 360K, respectively.} \label{fgr:fig4}
\end{figure}

The phase separation can be enhanced by the crystallisation of one of the blend components; here, the PCBM phase. By averaging the $S(Q,E)$ signal in energy, we can extract the neutron diffractograms of the samples, therefore allowing us to study miscibility and  crystallisation of PCBM simultaneously (Figure \ref{fgr:fig4}). The background in the diffractograms presented in Figures \ref{fgr:fig4} b, c and d is due to the incoherent contribution to the signal, and it follows the calculated trend (Figure \ref{fgr:fig1}). RRa-P3HT is indeed mainly amorphous with a broad Bragg peak around Q ~ 1.4 Å$^{-1}$, which in terms of distance is linked with the $\pi-\pi$ interaction and stacking in the material. As the PCBM concentration increases, peaks signature of PCBM crystallisation can be observed, where for instance the h-RRa-P3HT:h-PCBM 75wt\% h-PCBM sample is clearly crystalline. Although h-PCBM is highly crystalline at 360K, it is not easy to distinguish the crystalline signal in the d-RRa-P3HT:h-PCBM blends due to the small difference in the coherent cross sections of d-RRa-P3HT and h-PCBM. However, the presence of small peaks related to h-PCBM confirms that crystals of PCBM are present in the d-RRa-P3HT:h-PCBM blend with 50 wt\% h-PCBM.
\subsection{Probing changes in morphology by means of inelastic neutron scattering}
In a recent work,\cite{Guilbert2019} we used synergistically various neutron scattering techniques, including inelastic neutron scattering (INS), to map out the structural dynamics of RR-P3HT and RRa-P3HT up to the nanosecond time scale. Here, we use INS to probe the changes in morphology in the two phases present in the different blends of RRa-P3HT:PCBM: (i) an amorphous mixture of RRa-P3HT:PCBM with a concentration of about 20 wt\% PCBM for the blend h-RRa-P3HT:h-PCBM and about 30 wt\% for the blend d-RRa-P3HT:h-PCBM, and (ii) a PCBM-rich phase. So far, given that we considered mainly the incoherent scattering of the samples, we assumed that the PCBM-rich phase in those blends were similar to the neat PCBM samples. However, RRa-P3HT is likely to be also miscible to a lesser extent with PCBM and thus, the morphology of PCBM is likely to be different in the PCBM-rich phase compared to the neat PCBM phase.\\
\begin{figure}[H]
\includegraphics[width=18cm]{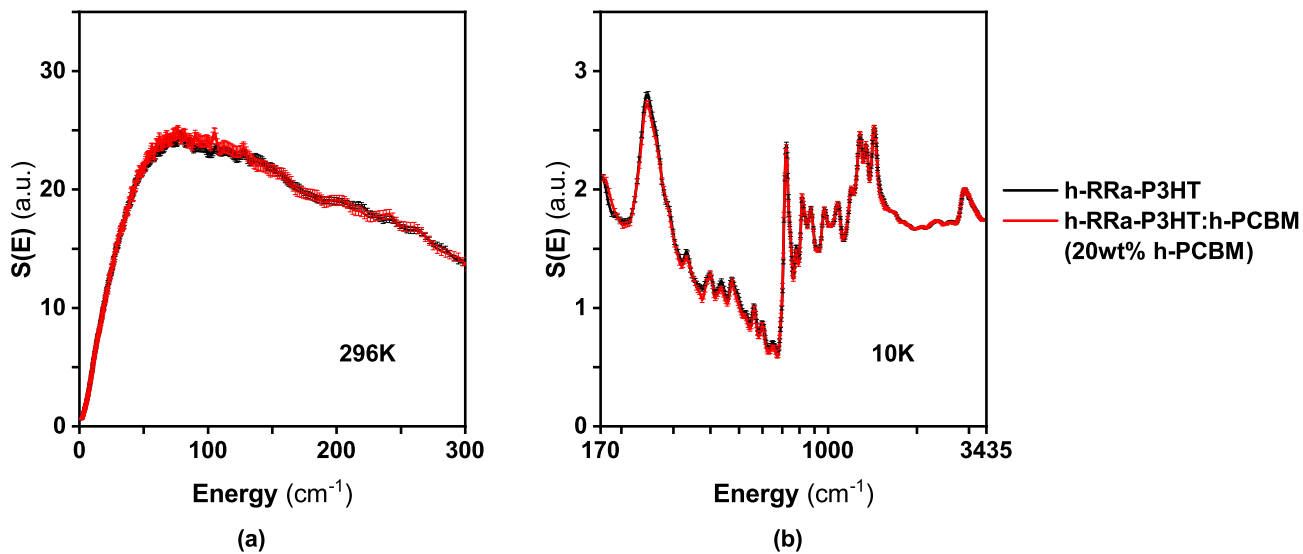}
\caption{INS spectra measured using (a) IN6 at 296K, and (b) IN1-Lagrange at 10 K of neat h-RRa-P3HT and blend h-RRa-P3HT:h-PCBM, with 20wt\% h-PCBM composition.} \label{fgr:fig5}
\end{figure}
Figure \ref{fgr:fig5} compares the INS spectra of the blend h-RRa-P3HT:h-PCBM, with 20 wt\% of h-PCBM (at the miscibility limit), and the neat sample h-RRa-P3HT. The full energy range is captured thanks to the combination of both the measurements shown in Figures \ref{fgr:fig5} a and b. No significant  differences are observed between the INS spectra of neat h-RRa-P3HT and the blend. Either no noticeable changes in morphology of h-RRa-P3HT occurs upon blending with h-PCBM, or the strong signal from hydrogens dominates the spectra and could therefore mask potential differences. Hydrogens are mainly located on the side-chains and therefore, their strong incoherent signal does not reflect strongly potential conformational changes or dynamics of the polymer backbones.
\begin{figure}[H]
\includegraphics[width=18cm]{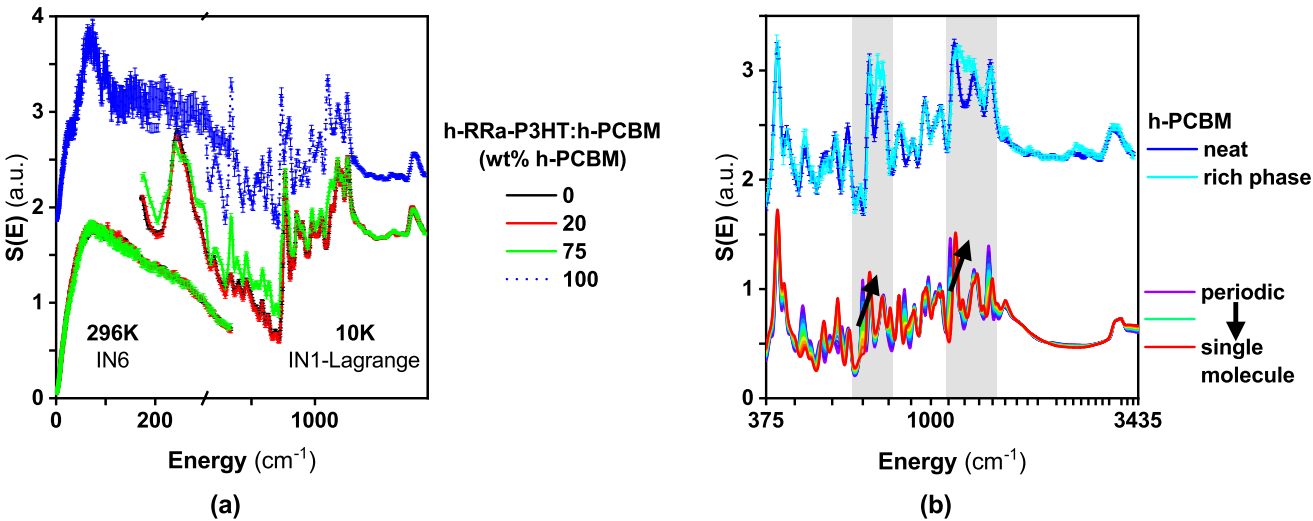}
\caption{(a) Composition-dependent INS spectra of the blend h-RRa-P3HT:h-PCBM measured using IN6 at 296K, and IN1-Lagrange at 10K, with variable h-PCBM concentrations of 0 wt\% (neat h-RRa-P3HT), 20 wt\% h-PCBM (at the miscibility limit), 75 wt\% h-PCBM and 100 wt\% (neat h-PCBM). (b) Comparison between measured and calculated INS spectra of h-PCBM. The measured INS spectra consist of neat h-PCBM phase and rich h-PCBM phase. The calculated INS spectra are from single molecule quantum chemical calculation and solid-state periodic calculation. The grey-shaded areas highlight regions with marked differences both in peak intensity and energy shift.} \label{fgr:fig6}
\end{figure}
As expected from our two-phase model, at a PCBM concentration (75 wt\% PCBM) much higher than the miscibility limit, $\mu$, the INS spectrum of the blend h-RRa-P3HT:h-PCBM exhibits clear PCBM-related features (Figure \ref{fgr:fig6} a). However, the INS spectrum of the 75 wt\% PCBM blend cannot be reconstructed by neutron weighting and combining the INS spectrum of h-RRa-P3HT:h-PCBM with a concentration close to the miscibility limit (20 wt\% PCBM), and the INS spectrum of neat h-PCBM. Therefore, we assign the differences between the neutron weighted average of the INS spectra and the measured spectrum to changes in morphology between the neat h-PCBM and the PCBM-rich phase. Thus, the INS spectrum of the PCBM-rich phase is extracted from the difference between the INS spectra of the h-RRa-P3HT:h-PCBM blends of compositions above the miscibility limit and the blend at the miscibility limit taking into account the neutron weights (Figure \ref{fgr:fig6} b). The main noticeable differences between the neat PCBM and the PCBM-rich phase are highlighted in the grey-shaded areas in Figure \ref{fgr:fig6} b. To get a deeper insight into these changes upon blending, we carried out density functional theory (DFT) calculations. Both the gas-phase single/isolated molecule and solid-state periodic approaches were adopted. This allows assessing (i) the relative strength of the intra-molecular and inter-molecular vibrational aspects and (ii) the importance of external (lattice) degrees-of-freedom. The single molecule approximation neglects the interaction with the environment, while the periodic approach accounts for the lattice dynamical modes and can point towards possible mode coupling between the molecular vibrations (internal modes) and the lattice (external) modes or between molecular modes. The INS spectrum of the neat h-PCBM is well simulated using the solid-state periodic approach, while the INS spectrum of the h-PCBM rich phase is found to be better approximated by the single molecule calculation. We conclude then that the crystallinity aspect of the neat PCBM is well described by the periodic calculations while the more disordered behaviour of the PCBM rich phase makes it reasonable to use the single molecule approximation. Since the focus is on an energy range where intra-molecular vibrations are dominating, we suggest that the coupling between intra-molecular vibrations and the environment decreases in the h-PCBM rich phase and any possible effects are potentially averaged out.
\begin{figure}[H]
\includegraphics[width=18cm]{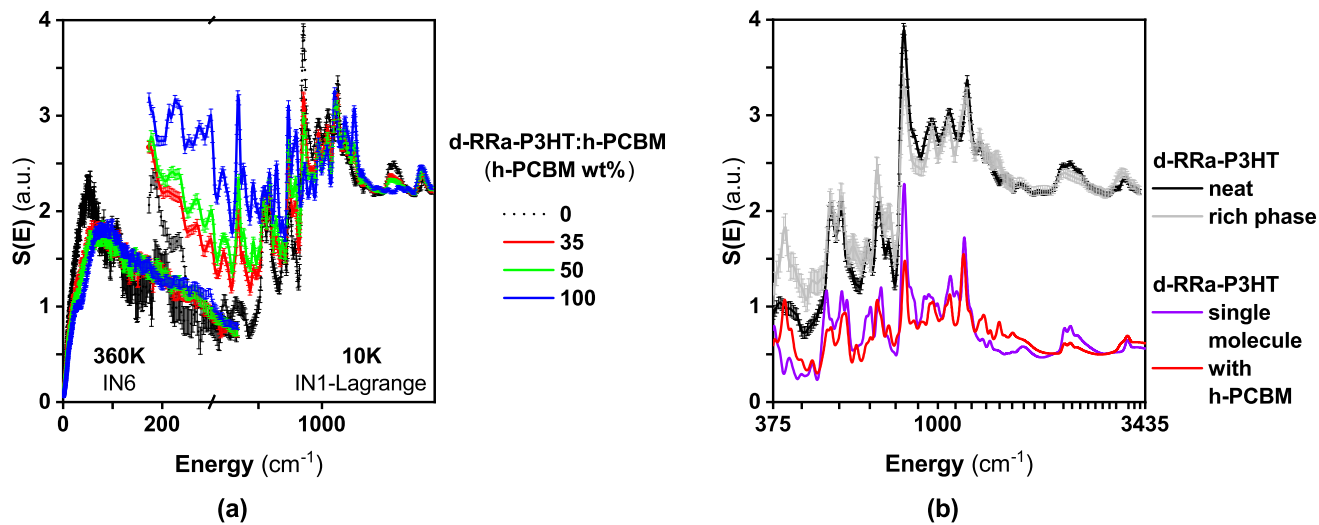}
\caption{(a) Composition-dependent INS spectra of the d-RRa-P3HT:h-PCBM blends measured using IN6 at 360K, and IN1-Lagrange at 10K, with variable h-PCBM concentrations of 0 wt\% (neat d-RRa-P3HT), 35 wt\% (close to the miscibility limit), 50 wt\% and 100 wt\% (neat h-PCBM). (b) Comparison between measured and calculated INS spectra of neat and rich d-RRa-P3HT phase. The measured INS spectra consist of neat d-RRa-P3HT phase and the reconstructed rich d-RRa-P3HT phase. The calculated INS spectra are from single-molecule quantum chemical calculations of a deuterated P3HT molecule and a deuterated P3HT molecule with a PCBM molecule.} \label{fgr:fig7}
\end{figure}
In order to resolve any possible changes of conformation or coupling between P3HT and PCBM in the mixed phase, we measured the partially deuterated blends d-RRa-P3HT:h-PCBM with 35 wt\% and 50 wt\% of h-PCBM (Figure \ref{fgr:fig7} a). Note that 35 wt\% h-PCBM is close to the miscibility limit of 30 wt\% h-PCBM, determined above. Clear differences are observed between INS spectra of neat d-RRa-P3HT and d-RRa-P3HT:h-PCBM with 35 wt\% h-PCBM, and are mainly assigned to features related to h-PCBM. These differences become marked by increasing the h-PCBM concentration. We went a step further by using the INS spectrum of the h-PCBM rich phase to reconstruct the d-RRa-P3HT rich phase spectrum (Figure \ref{fgr:fig7} b). The INS spectrum of the d-RRa-P3HT-rich phase is extracted as the difference between the INS spectra of d-RRa-P3HT:h-PCBM blends of compositions above the miscibility limit and the h-PCBM rich phase taking into account the neutron weights. The DFT-based single oligomer calculation of d-RRa-P3HT reproduced well the INS spectrum of neat d-RRa-P3HT, which is mainly amorphous. Note that d-RRa-P3HT has hydrogenated defects along the backbones and the strong peaks at 815 cm$^{-1}$ and 1185 cm$^{-1}$ can only be reproduced by adding hydrogen defects along the backbone in the model calculations. The observed differences between the measured INS spectra of the neat d-RRa-P3HT and rich d-RRa-P3HT phase are captured by our reconstruction method consisting of neutron weighting and combining the single molecule calculation for d-RRa-P3HT and the single molecule calculation for h-PCBM. No noticeable changes in morphology of RRa-P3HT or coupling of modes between RRa-P3HT and PCBM are observed.

\section{Experimental}
h-RRa-P3HT was obtained from Sigma-Aldrich. d-RRa-P3HT was synthesized by an iron(III) chloride mediated oxidative polymerization of 4-d1-3-d13-hexylthiophene in chloroform at room temperature. The molecular weights, polydispersities, and regioregularities are summarized in Table \ref{tbl:tbl_1}. The polymers in this study come from the same batches as our previous study\cite{Guilbert2019}. PCBM $>$ 99\% grade was obtained from Solenne BV. The as-received materials were dissolved in chloroform (40 mg/mL) and drop-cast on a glass slide on a hot plate at 60 °C for 1 h. The drop-cast films were then scratched from the glass substrates and stacked in aluminum foil. Each measured sample was about 400 mg. Further details related to the materials and their characterisation can be found in references. \cite{Guilbert2016, Guilbert2019}.\\
\begin{table}[h]
\begin{center}
\caption{Molecular weight (Mw) and polydispersity (PDI) as measured by gel permeation chromatography, regioregularity (RR) as measured by NMR and scattering length density (SLD) calculated assuming a density of 1 g/cm$^{3}$.}
\label{tbl:tbl_1}
\begin{tabular}{ccccc}
\hline
 & Mw (in kDa) & PDI & RR (\%) & SLD (in 10$^{-6}$ \r{A}$^{-2}$)\\
\hline
h-RRa-P3HT & 304 & 3.2 & 56 & 0.61417\\
d-RRa-P3HT &  53 & 1.95 & 73 & 5.4053\\
\hline
\end{tabular}
\end{center}
\end{table}
\\The neutron spectroscopy measurements were performed using the direct geometry, cold neutron, time-of-flight, time-focusing spectrometer IN6, and the hot-neutron, inverted geometry spectrometer IN1-Lagrange at the Institut Laue-Langevin (ILL, Grenoble, France). Data were reduced, treated and analysed in a similar way as was done in our previous related works \cite{Guilbert2016, Guilbert2019}.\\
DFT-based quantum chemical isolated molecule and solid-state periodic calculations were performed using Gaussian 16 \cite{g16} and Castep \cite{castep}, respectively. For the isolated molecules, the functional/basis-set b3lyp/6-311g(d,p) was chosen \cite{b3lyp}. For the solid-state periodic calculations, the functional PBE \cite{pbe} with van der Waals corrections \cite{vdw} were used. Full computational details can be found in reference \cite{Guilbert2019}.

\section{Conclusions}
We presented a neutron spectroscopy based methodology to study phase behavior and morphology of the blend system P3HT:PCBM. We used a variable PCBM composition approach and deuteration technique for P3HT to determine the miscibility limits of the fullerene within the regio-random (amorphous) form of P3HT (RRa-P3HT). Temperature-dependent and composition-dependent quasi-elastic neutron scattering and inelastic neutron scattering measurements were performed to evaluate the phase composition and behaviour of the blends, to monitor simultaneously their miscibility and microstructure evolution and to probe changes in their morphology. This approach enabled us to resolve the evolution of the microstructure and morphology that are correlated with changes in structural dynamics of the polymer and fullerene upon blending. Our approach using single-molecule and solid-state periodic DFT calculations could reproduce the differences in INS spectra between crystalline neat h-PCBM and the more disordered h-PCBM rich phase. However, no clear evidences of P3HT conformational changes over blending could be concluded. It should be reminded though that neutrons probe an ensemble of conformations of both neat d-RRa-P3HT phase and d-RRa-P3HT rich phase. Therefore, our approach might be limited in the capture of morphological changes that will affect only chains that are in close contact with the PCBM.

\section*{Acknowledgements}
The ILL is acknowledged for the use of the spectrometers IN1-Lagrange and IN6. A. A. Y. G. acknowledges EPSRC for the award of an EPSRC Postdoctoral Fellowship (EP/P00928X/1).
\section*{Author contributions}
A. A. Y. G. and M. Z. conceived and developed the neutron-based project, prepared the samples, performed the neutron measurements, treated, analysed and interpreted the neutron data and wrote the manuscript. P. A. F. and C. B. N. synthesised d-RRa-P3HT.
\providecommand*{\mcitethebibliography}{\thebibliography}
\csname @ifundefined\endcsname{endmcitethebibliography}
{\let\endmcitethebibliography\endthebibliography}{}

\end{document}